\documentclass[11pt]{article}
\usepackage{amsmath,amssymb,graphicx}

\newcommand{\ba}{\begin{alignat}{3}}

\newcommand{\dl}{\delta}

\newcommand{\pa}{\partial}

\newcommand{\om}{\omega}

\begin{document}

\begin{titlepage}
\begin{flushright}
\end{flushright}
\begin{center}
  \vspace{4cm}
  {\bf \Large Quantum M-wave and Black 0-brane}
  \\  \vspace{2cm}
  Yoshifumi Hyakutake
   \\ \vspace{1cm}
   {\it Faculty of Science, Ibaraki University \\
   Bunkyo 2-1-1, Mito, Ibaraki 310-8512, Japan}
\end{center}

\vspace{2cm}
\begin{abstract}
The effective action of superstring theory or M-theory is approximated by supergravity in the low energy limit,
and quantum corrections to the supergravity are taken into account by including higher derivative terms. 
In this paper, we consider equations of motion with those higher derivative terms in M-theory and
solve them to derive quantum M-wave solution. 
A quantum black 0-brane solution is also obtained by Kaluza-Klein dimensional reduction of 
the M-wave solution.
The quantum black 0-brane is asymptotically flat and uniquely determined by imposing appropriate conditions.
The mass and the R-R charge of the quantum black 0-brane are derived by using the ADM mass and 
the charge formulae, and we see that only the mass is affected by the quantum correction.
Various limits of the quantum black 0-brane are also considered, and especially
we show that an internal energy in the near horizon limit is correctly reproduced.
\end{abstract}
\end{titlepage}

\setlength{\baselineskip}{0.65cm}

\section{Introduction}

Superstring theory is a promising candidate for the theory of quantum gravity.
UV divergence of the gravity is well controlled and it unifies gauge theory and gravity consistently.
Therefore much effort has been devoted to reveal the quantum nature of the gravity.
In this paper, we proceed those discussions and describe the quantum geometry in the superstring theory.

The superstring theory contains not only fundamental strings but also D$p$-branes 
which extend $p$ spacial directions\cite{Pol}.
In the low energy limit, the superstring theory is well approximated by the supergravity, and
D$p$-branes are described by a classical solution which is called a black $p$-brane\cite{GM,HS}. 
In the extremal case where the mass and Ramond-Ramond charge of the black $p$-brane are balanced,
quantum corrections to the classical solution are suppressed. 
Then for a special class of black brane solutions, a statistical derivation of its entropy is possible 
from the dual gauge theory on the corresponding D-branes\cite{SV}.
Furthermore, by taking a near horizon limit of extremal black branes the geometry 
becomes anti-de-Sitter space-time, and it is conjectured that
the superstring theory in the AdS background 
is dual to the gauge theory on the D-branes\cite{M}. 
Correlation functions of the gauge theory can be calculated from the dual gravity theory\cite{GKP,W2}.
On the other hand, the entropy counting or test of the gauge gravity duality 
of non-extremal black branes is quite difficult because quantum corrections become important 
which are not well understood so far. 
In this paper, we discuss quantum corrections to the non-extremal black 0-brane by explicitly solving
equations of motion\footnote{Higher derivative corrections are also important
for the extremal case\cite{D,OSV}.}.

In order to investigate quantum nature of the non-extremal black $p$-brane, we need to know
quantum corrections to the supergravity.
As the superstring theory is defined perturbatively, it is possible to derive
corrections to the supergravity so as to be consistent with the scattering amplitude\cite{GW,GS}
or $\sigma$-model calculations\cite{GVZ,GZ}.
Among these corrections, the structure of higher curvature $R^4$ terms is well studied\cite{Ts,BB}
and they come from 4 gravitons amplitudes at tree and 1-loop levels\cite{GW,GS,PT}. 
$R^4$ terms are also derived by imposing local supersymmetry\cite{RSW1}-\cite{H}.
There are several attempts to solve the modified equations of motion\cite{CMP}-\cite{B}.
For the black $p$-brane, however, it is difficult to solve the equations of motion consistently
since the full form of the effective action including R-R gauge fields is not completely determined 
so far\footnote{In three dimensions the higher derivative corrections are well controlled.
Although physical quantities are affected by the quantum corrections, 
the solution is the same as the classical one\cite{SS}.}.
In this paper we concentrate on the black 0-brane in type IIA supergravity, 
thus at least the knowledge of quantum corrections to a metric, a dilaton field and 
a R-R 1-form filed are necessary. 
This problem is resolved by noting that these fields are gathered into the metric in 11 dimensions\cite{H2}.
Fortunately supersymmetric higher curvature corrections to the 11 dimensional supergravity 
are well known\cite{PVW,HO,H}, so usual Kaluza-Klein dimensional reduction gives 
1-loop quantum corrections to the type IIA supergravity which are relevant to the black 0-brane.

By taking account of the $R^4$ terms, quantum corrections to the near horizon geometry of 
the non-extremal black 0-brane is analytically solved in ref.~\cite{H2}. 
This is quite useful to test the gauge gravity duality, since
computer simulations of the dual gauge theory on the D0-branes are well developed in refs.~\cite{KLL}-\cite{N}.
In fact, the test of the gauge gravity duality is examined in ref.~\cite{HHIN}, including 
$1/N^2$ quantum corrections to the classical gravity. 
And the analytic result from the gravity side 
is well reproduced from the gauge theory side numerically\cite{HHIN}. 
See also ref.~\cite{Hanada} for introductory review.

In this paper, we generalize the discussions in ref.~\cite{H2} to derive quantum corrections to  
asymptotically flat non-extremal M-wave and black 0-brane.
We often call these solutions quantum M-wave and quantum black 0-brane.
Organization of this paper is as follows.
In section 2, we briefly review the classical M-wave and black 0-brane solution.
The main part of this paper appears in section 3, where the quantum corrections to the non-extremal
M-wave and black 0-brane are analytically solved. 
In section 4, we consider extremal limit, Schwarzschild one
and near horizon one for the quantum black 0-brane. 
In section 5, the ADM mass and the R-R charge of the quantum black 0-brane are derived.
Section 6 is devoted to conclusion and discussion.
In the appendix, we employ Noether and Wald's method to derive the quantum corrections 
to the ADM mass and the charge formulae.
We also discuss the near horizon limit of the internal energy of the quantum black 0-brane.

\section{Review of Classical M-wave and Black 0-brane}

It is conjectured that the strong coupling limit of type IIA superstring theory in 10 dimensions 
is described by M-theory in 11 dimensions\cite{T,W3}. 
The string coupling constant $g_s$ and string length $\ell_s$ in the type IIA superstring theory 
are related to the radius of 11th circle $R_{11}=g_s\ell_s$ and 11 dimensional Planck length 
$\ell_p = g_s^{1/3} \ell_s$ in the M-theory.
A D0-brane is identified with a Kaluza-Klein mode, and the mass of the D0-brane is given by $1/R_{11}$.
In the low energy limit, the type IIA superstring theory and M-theory are approximated by
type IIA supergravity and 11 dimensional supergravity, respectively.
Then the D0-brane is approximated by a black 0-brane solution in the type IIA supergravity,
and the Kaluza-Klein mode is described by a M-wave solution in the 11 dimensional supergravity.
In this section, we briefly review classical properties of the black 0-brane via dimensional reduction
of the M-wave solution. 
Simultaneously we fix the notations and conventions used in this paper.

The fields of the 11 dimensional supergravity consist of a graviton $g_{MN}$, 
a 3-form field $A_{MNP}$ and a Majorana gravitino $\psi_M$~\cite{CJS}.
Here $M,N,P$ are space-time indices in 11 dimensions, and both bosonic and fermionic fields
have 128 physical degrees of freedom.
Since we only consider the M-wave solution which does not couple to $A_{MNP}$ and $\psi_M$,
a relevant part of the action is simply given by
\begin{alignat}{3}
  S_{11}^{(0)} &= \frac{1}{2\kappa_{11}^2} \int d^{11}x\, \sqrt{-g} R, \label{eq:11act}
\end{alignat}
where $2\kappa_{11}^2 = (2\pi)^8 \ell_p^9$.
The equations of motion become $R_{MN}- \frac{1}{2} g_{MN} R = 0$, 
and the following geometry becomes a solution.
\begin{alignat}{3}
  &ds_{11}^2 = - H^{-1} F dt^2 + F^{-1} dr^2 + r^2 d\Omega_8^2 
  + \Big( H^{\frac{1}{2}} dz - \Big(\frac{r_+}{r_-}\Big)^\frac{7}{2} H^{-\frac{1}{2}} dt \Big)^2,
  \label{eq:11sol}
  \\
  &H = 1 + \frac{r_-^7}{r^7}, \qquad F = 1 - \frac{r_+^7 - r_-^7}{r^7}. \notag
\end{alignat}
This is the non-extremal M-wave solution, which contains two parameters $r_\pm$.
The extremal case is saturated when $r_+ = r_-$, and Schwarzschild black hole smeared along $z$ direction
is obtained when $r_- \to 0$.

The type IIA supergravity consists of a graviton $g_{\mu\nu}$, a dilaton $\phi$, a R-R 1-form field $C_\mu$, 
a NS-NS 2-form field $B_{\mu\nu}$, a R-R 3-form field $C_{\mu\nu\rho}$, a Majorana gravitino 
$\psi_\mu$ and a Majorana dilatino $\psi$.
Here $\mu,\nu,\rho$ are space-time indices in 10 dimensions, and both bosonic and fermionic fields
have the same physical degrees of freedom as those in 11 dimensional supergravity.
The type IIA supergravity is derived by dimensional reduction of the 11 dimensional supergravity
if we express the metric in 11 dimensions as
\begin{alignat}{3}
  g_{MN} dx^M dx^N = e^{-2\phi/3} g_{\mu\nu} dx^\mu dx^\nu + e^{4\phi/3} (dz - C_\mu dx^\mu)^2,
  \label{eq:dimred}
\end{alignat}
where $z$ direction is the circle with the radius $R_{11}$\cite{HN}.
By inserting this metric into the action~(\ref{eq:11act}), we obtain the 10 dimensional action of the form
\ba
  S_{10}^{(0)} &= \frac{1}{2\kappa_{10}^2} \int d^{10} x \sqrt{-g} 
  \Big\{ e^{-2\phi} \big( R + 4 \partial_\mu \phi \partial^\mu \phi \big) 
  - \frac{1}{4} G_{\mu\nu} G^{\mu\nu} \Big\},
  \label{eq:10dsugra}
\end{alignat}
where $2\kappa_{10}^2 = (2\pi)^7 \ell_s^8 g_s^2$ and $G_{\mu\nu}$ is the field strength of $C_\mu$.
Notice that the coordinate transformation $z' = z + \chi(x)$ in 11 dimensions corresponds to 
the gauge transformation $C'_\mu = C_\mu - \partial_\mu \chi$ in 10 dimensions.

The non-extremal black 0-brane solution is obtained by the dimensional reduction of the 
M-wave solution (\ref{eq:11sol}). By applying eq.~(\ref{eq:dimred}), we obtain\cite{AGMOO}
\begin{alignat}{3}
  &ds_{10}^2 = - H^{-\frac{1}{2}} F dt^2 + H^{\frac{1}{2}} F^{-1} dr^2 
  + H^\frac{1}{2} r^2 d\Omega_8^2 , \quad
  e^\phi = H^\frac{3}{4}, 
  \quad C = \Big(\frac{r_+}{r_-}\Big)^\frac{7}{2} H^{-1} dt. \label{eq:10sol}
\end{alignat}
Thus the purely geometrical object in 11 dimensions becomes the charged black hole in 10 dimensions.
The event horizon is located at 
\begin{alignat}{3}
  r_H = (r_+^7 - r_-^7)^{\frac{1}{7}} \equiv r_- \alpha, \label{eq:alpha}
\end{alignat}
where $\alpha$ is a dimensionless parameter.
And the mass $M$ and the R-R charge $Q$ of the black 0-brane are evaluated as
\begin{alignat}{3}
  M = \frac{V_{S^8}}{2\kappa_{10}^2} \big( 8 r_+^7 - r_-^7 \big), \qquad
  Q = \frac{V_{S^8}}{2\kappa_{10}^2} 7 \big( r_+ r_- \big)^{\frac{7}{2}}. \label{eq:MQ}
\end{alignat}
$V_{S^8} = \frac{2 \pi^{9/2}}{\Gamma(9/2)} = \frac{2(2\pi)^4}{7\cdot 15}$ is the volume of $S^8$.
Since the charge of $N$ D0-branes is quantized as $Q = \frac{N}{\ell_s g_s}$ in the type IIA superstring theory, 
the parameters $r_\pm$ can be expressed as
\begin{alignat}{3}
  r_\pm^7 &= (1+\delta)^{\pm 1}(2\pi)^2 15\pi g_s N \ell_s^7, \label{eq:rpm}
\end{alignat}
by introducing a non-negative parameter $\delta$. 
There are three limits of the solution (\ref{eq:10sol}) which are important in later sections.
(See table.~\ref{table:limits}.)
First one is the extremal limit $r_+ \to r_-$ which is equivalent to $\alpha \to 0$.
Second one is the Schwarzschild limit $r_- \to 0$, where the charge $Q$ goes to zero.
Final one is the near horizon limit which is realized by $r \to 0$ with $U \equiv r/\ell_s^2$, 
$\lambda \equiv g_s N/(2\pi)^2\ell_s^3$ and $U_0 \equiv r_-\alpha/\ell_s^2$ fixed\cite{IMSY}.
This is equivalent to $\alpha \to 0$ by fixing $r_-\alpha/r$, $\ell_s^2/(r_-\alpha)$
and $g_s^2 N^2/(r_- \alpha)^3$, so the near horizon limit corresponds to the near extremal limit.
\\

\begin{table}[htb]
\begin{center}
\begin{tabular}{l|c}
  Extremal limit & $\alpha \;\to\; 0$ \\\hline
  Schwarzschild limit & $r_- \to 0$ \\\hline
  Near horizon limit & $\alpha \;\to\; 0$ by fixing $\frac{r_- \alpha}{r}$, $\frac{\ell_s^2}{r_-\alpha}$ 
  and $\frac{g_s^2 N^2}{(r_- \alpha)^3}$
\end{tabular}
\caption{Extremal, Schwarzschild and near horizon limits} \label{table:limits}
\end{center}
\end{table}

\vspace{-0.5cm}
\section{Quantum M-wave and Black 0-brane}\label{sec:solve}

In this section, we consider leading quantum correction to the M-wave solution (\ref{eq:11sol}).
Since the M-wave solution is purely geometrical in 11 dimensions, the 3-form $A_{MNP}$
is irrelevant to our analyses.
Thus it is enough to investigate the effective action of the M-theory which depends only 
on higher curvature terms.
The leading structure of those is known to be $R^4$ terms\cite{Ts,BB}.
The explicit form of the effective action is given by
\begin{alignat}{3}
  S_{11} &= \frac{1}{2 \kappa_{11}^2} \int d^{11}x \; e \Big\{ R +
  \gamma \Big(t_8 t_8 R^4 - \frac{1}{4!} \epsilon_{11} \epsilon_{11} R^4 \Big) \Big\}
  \notag
  \\
  &= \frac{1}{2 \kappa_{11}^2} \int d^{11}x \; e \Big\{ R +
  24 \gamma \big( R_{abcd} R_{abcd} R_{efgh} R_{efgh}
  - 64 R_{abcd} R_{aefg} R_{bcdh} R_{efgh} \notag
  \\
  &\qquad\qquad\qquad\qquad\qquad\quad
  + 2 R_{abcd} R_{abef} R_{cdgh} R_{efgh}
  + 16 R_{acbd} R_{aebf} R_{cgdh} R_{egfh} \notag
  \\
  &\qquad\qquad\qquad\qquad\qquad\quad
  - 16 R_{abcd} R_{aefg} R_{befh} R_{cdgh}
  - 16 R_{abcd} R_{aefg} R_{bfeh} R_{cdgh} \big)
  \Big\}. \label{eq:R4}
\end{alignat}
Although we neglected fermionic terms, a part of them is also obtained in refs.~\cite{PVW,HO,H}.
The expansion parameter in the action is given by
\begin{alignat}{3}
  \gamma = \frac{\pi^2\ell_p^6}{2^{11} 3^2},
\end{alignat} 
and $a,b,c,\cdots = 0,1,\cdots,10$ are local Lorentz indices.
All indices are lowered for simplicity but should be contracted by the flat metric.
Note that $\gamma \sim g_s^2 \ell_s^6$, so when the effective action (\ref{eq:R4}) is reduced 
to ten dimensions, it becomes one-loop leading corrections to the type IIA supergravity.
By varying the effective action (\ref{eq:R4}), equations of motion are obtained as
\begin{alignat}{3}
  E_{ij} &\equiv R_{ij} - \frac{1}{2} \eta_{ij} R + \gamma \Big\{
  - \frac{1}{2} \eta_{ij} \Big(t_8 t_8 R^4 - \frac{1}{4!} \epsilon_{11} \epsilon_{11} R^4 \Big) \notag
  \\
  &\quad
  + \frac{3}{2} R_{abci} X^{abc}{}_j - \frac{1}{2} R_{abcj}X^{abc}{}_i - 2 D_{(a} D_{b)} X^a{}_{ij}{}^b 
  \Big\} = 0. \label{eq:MEOM}
\end{alignat}
Here $D_a$ is a covariant derivative for local Lorentz indices and
\begin{alignat}{3}
  X_{abcd} &= \frac{1}{2} \big( X'_{[ab][cd]} + X'_{[cd][ab]} \big), \label{eq:X}
  \\
  X'_{abcd} &= 96 \big(
  R_{abcd} R_{efgh} R_{efgh} - 16 R_{abce} R_{dfgh} R_{efgh} + 2 R_{abef} R_{cdgh} R_{efgh} \notag
  \\
  &\qquad\,
  + 16 R_{aecg} R_{bfdh} R_{efgh} - 16 R_{abeg} R_{cfeh} R_{dfgh} - 16 R_{efag} R_{efch} R_{gbhd} \notag
  \\
  &\qquad\,
  + 8 R_{abef} R_{cegh} R_{dfgh} \big) \notag.
\end{alignat}
The details of the derivation can be found in ref.~\cite{H2}.

Let us solve the eq.~(\ref{eq:MEOM}) up to the linear order of $\gamma$.
The leading part of the metric (\ref{eq:11sol}) itself is not a solution of the eq.~(\ref{eq:MEOM}),
we should relax the ansatz for the M-wave. 
Most general static ansatz with SO(9) rotation symmetry is given by
\begin{alignat}{3}
  ds_{11}^2 &= - H_1^{-1} F_1 dt^2 + F_1^{-1} dr^2
  + r^2 d\Omega_8^2 + \big( H_2^{\frac{1}{2}} dz 
  - \sqrt{1+\alpha^7} H_3^{-\frac{1}{2}} dt \big)^2, \label{eq:Mansatz}
  \\
  H_i &= 1 + \frac{r_-^7}{r^7} + \frac{\gamma}{r_-^6 \alpha^{13}} h_i \Big(\frac{r}{r_-\alpha}\Big), \qquad
  F_1 = 1 - \frac{r_-^7 \alpha^7}{r^7} + \frac{\gamma}{r_-^6 \alpha^6} f_1 \Big(\frac{r}{r_-\alpha}\Big). \notag
\end{alignat}
Here $\alpha$ is given by $\alpha = (r_+^7/r_-^7 - 1)^\frac{1}{7}$,
and $h_i (i=1,2,3)$ and $f_1$ are functions of $\frac{r}{r_-\alpha}$.
Note that, up to the linear order of $\gamma$, 
the coordinate transformation $dz \to dz + c_g dt$ is interpreted as the change of $h_2(x)+h_3(x)$,
\begin{alignat}{3}
  \frac{\gamma}{r_-^6 \alpha^{13}} (h_2 + h_3) \to 
  \frac{\gamma}{r_-^6 \alpha^{13}} (h_2 + h_3) 
  + \frac{2c_g}{\sqrt{1+\alpha^2}} \Big( 1 + \frac{r_-^7}{r^7} \Big)^2. \label{eq:gaugetr}
\end{alignat}
It is clear that this corresponds to the gauge transformation of $C_\mu$ in 10 dimensions.

Now we insert the ansatz (\ref{eq:Mansatz}) into the eq.~(\ref{eq:MEOM}).
In order to make the equations of motion simple, we introduce following dimensionless coordinates,
\begin{alignat}{3}
  \tau = \frac{t}{r_-\alpha}, \qquad x = \frac{r}{r_-\alpha}, \qquad y = \frac{z}{r_-\alpha}.
\end{alignat}
Then all components of the metric are expressed as a function of $x$, and
we obtain following 5 differential equations for $h_i(x)$ and $f_1(x)$\footnote{
In order to derive the equations of motion, 3 independent Mathematica codes are used.}.
\begin{alignat}{3}
  E_1 &= 
  - 7 ( 9 + 32 \alpha^7 x^{7} + 16 \alpha^{14} x^{14} )( 1 + \alpha^7 x^7 ) x^{34} f_1 
  - ( 9 + 16 \alpha^7 x^{7} ) ( 1 + \alpha^7 x^7 )^2 x^{35} f_1' \notag
  \\
  &\quad\,
  - 49 (1 + \alpha^7) x^{41} h_1
  + 49 (1 - x^7) x^{34} h_2
  + (1 - x^7) ( 23 + 16 \alpha^7 x^{7} ) ( 1 + \alpha^7 x^7 ) x^{35} h_2' \notag
  \\
  &\quad\,
  + 2 (1 - x^7) ( 1 + \alpha^7 x^7 )^2 x^{36} h_2'' 
  + 98 ( 1 + \alpha^7 ) x^{41} h_3 
  + 7 (1 + \alpha^7) ( 1 + \alpha^7 x^7) x^{42} h_3' \label{eq:1}
  \\
  &\quad\,
  + 17418240 ( 61 + (4032 + 4093 \alpha^7) x^7 - 3640 (1+ \alpha^7) x^{14} ) ( 1 + \alpha^7 x^7 )^2 = 0, 
  \notag
  \\[0.2cm]
  E_2 &= 
  7 ( 9 + 32 \alpha^7 x^{7} + 16 \alpha^{14} x^{14} ) ( 1 + \alpha^7 x^7 ) x^{34} f_1 
  + ( 9 + 16 \alpha^7 x^{7} ) ( 1 + \alpha^7 x^7 )^2 x^{35} f_1' \notag
  \\
  &\quad\,
  + 7 ( 9 - (2 - 23 \alpha^7) x^7 - 16 \alpha^7 x^{14} ) x^{34} h_1
  + ( 1 - x^7 ) ( 9 + 16 \alpha^7 x^{7} ) ( 1 + \alpha^7 x^7 ) x^{35} h_1'  \notag
  \\
  &\quad\,
  - 112 ( 1 - x^7 ) ( 1 + \alpha^7 x^{7} ) x^{34} h_2 
  - 16 ( 1 - x^7 ) ( 1 + \alpha^7 x^{7} )^2 x^{35} h_2'
  - 98 ( 1 + \alpha^7 ) x^{41} h_3 \label{eq:2}
  \\
  &\quad\,
  - 7 ( 1 + \alpha^7 ) ( 1 + \alpha^7 x^{7} ) x^{42} h_3' 
  - 17418240 \alpha^{34} ( 329 + 124 x^7 ) ( 1 + \alpha^7 x^7 )^3 = 0, \notag
  \\[0.2cm]
  E_3 &= 
  7 ( 19 + 24 \alpha^7 x^{7} + 12 \alpha^{14} x^{14} ) ( 1 + \alpha^7 x^{7} ) x^{34} f_1
  + 7 ( 5 + 4 \alpha^7 x^{7} ) ( 1 + \alpha^7 x^{7} )^2 x^{35} f_1' \notag
  \\
  &\quad\,
  + 2 ( 1 + \alpha^7 x^{7} )^3 x^{36} f_1''
  + 14 ( 6 - 5 (4 + 3 \alpha^7) x^7 + \alpha^7 x^{14} ) x^{34} h_1 \notag
  \\
  &\quad\,
  + 7 ( 4 - (7 + \alpha^7) x^7 - 2 \alpha^7 x^{14} ) ( 1 + \alpha^7 x^7 ) x^{35} h_1'
  + 2 ( 1 - x^7 ) ( 1 + \alpha^7 x^{7} )^2 x^{36} h_1'' \label{eq:3}
  \\
  &\quad\,
  - 7 ( 5 - (26 + 23 \alpha^7) x^7 + 2 \alpha^7 x^{14} ) x^{34} h_2
  - 7 ( 3 - (6 + \alpha^7) x^7 - 2 \alpha^7 x^{14} ) ( 1 + \alpha^7 x^7 ) x^{35} h_2' \notag
  \\
  &\quad\,
  - 2 (1 - x^7) ( 1 + \alpha^7 x^{7} )^2 x^{36} h_2''
  + 98 ( 1 + \alpha^7 ) x^{41} h_3
  + 7 ( 1 + \alpha^7 ) ( 1 + \alpha^7 x^{7} ) x^{42} h_3' \notag
  \\
  &\quad\,
  - 30481920 \alpha^{34} ( 283 - 186 x^7 ) ( 1 + \alpha^7 x^7 )^3 = 0, \notag
  \\[0.2cm]
  E_4 &= 
  7 ( 37 + 32 \alpha^7 x^{7} + 16 \alpha^{14} x^{14} ) ( 1 + \alpha^7 x^{7} ) x^{34} f_1
  + ( 53 + 32 \alpha^7 x^{7} ) ( 1 + \alpha^7 x^{7} )^2 x^{35} f_1' \notag
  \\
  &\quad\,
  + 2 ( 1 + \alpha^7 x^7 )^3 x^{36} f_1''
  + 147 ( 1 - (3 + 2 \alpha^7) x^7 ) x^{34} h_1 \notag
  \\
  &\quad\,
  + ( 37 - (58 + 5 \alpha^7) x^7 - 16 \alpha^7 x^{14} ) ( 1 + \alpha^7 x^7 ) x^{35} h_1'
  + 2 ( 1 - x^7 ) ( 1 + \alpha^7 x^{7} )^2 x^{36} h_1'' \notag
  \\
  &\quad\,
  + 147 ( 1 + \alpha^7 ) x^{41} h_2
  + 21 ( 1 + \alpha^7 ) ( 1 + \alpha^7 x^{7} ) x^{42} h_2' \label{eq:4}
  \\
  &\quad\,
  + 294 ( 1 + \alpha^7 ) x^{41} h_3
  + 21 ( 1 + \alpha^7 ) ( 1 + \alpha^7 x^{7} ) x^{42} h_3' \notag
  \\
  &\quad\,
  - 17418240 ( 4093 - (7672 - 61 \alpha^7) x^7 + 3640 x^{14} ) ( 1 + \alpha^7 x^7 )^2 = 0, \notag
  \\[0.2cm]
  E_5 &= 
  49 x^{34} h_1
  + 7 ( 1 + \alpha^7 x^7 ) x^{35} h_1' 
  + 49 x^{34} h_2
  - ( 1 + 8 \alpha^7 x^{7} ) ( 1 + \alpha^7 x^{7} ) x^{35} h_2' \notag
  \\
  &\quad\,
  - ( 1 + \alpha^7 x^{7} )^2 x^{36} h_2''
  - 98 x^{34} h_3
  - 2 ( 11 + 15 \alpha^7 x^{7} + 4 \alpha^{14} x^{14} ) x^{35} h_3' \label{eq:5}
  \\
  &\quad\,
  - ( 1 + \alpha^7 x^7 )^2 x^{36} h_3''
  + 975421440 \alpha^{27} ( 72 - 65 x^7 ) ( 1 + \alpha^7 x^7 )^2 = 0. \notag
\end{alignat}
The above equations come from the linear order of $\gamma$, and equations of
$\mathcal{O}(\gamma^0)$ are automatically satisfied.
In following subsections, we will solve 5 equations of motion step by step and determine $h_i(x)$ and $f_1(x)$.
Although there appear several integral constants, these are uniquely fixed by imposing appropriate conditions.
For example, we require that $h_i(x)$ and $f_1(x)$ do not diverge around the event horizon, and
$h_3(x)$ should be determined up to the gauge transformation (\ref{eq:gaugetr}).
Consistency conditions with the extremal limit and the Schwarzschild one in table \ref{table:limits} 
are also taken into account.
A complete form of the solution is summarized in section \ref{sec:sol}.

\subsection{Solve $E_1 - E_2 + E_3 - E_4 = 0$}

First let us consider a combination of $E_1 - E_2 + E_3 - E_4$.
After some calculations, it is simplified as
\begin{alignat}{3}
  0 &= \frac{E_1 - E_2 + E_3 - E_4}{36x^{28}(1+\alpha^7 x^7)^3} \notag
  \\
  &= - (x^7 f_1)' - \frac{1 - x^7}{2} 
  \Big\{ \frac{x^7 ( h_1 - h_2 )}{1 + \alpha^7 x^7} \Big\}' 
  + \frac{362880}{x^{28}} ( 5317 - 4254 x^7 ). 
\end{alignat}
From this equation, $(x^7 f_1)'$ is expressed in terms of $(h_1-h_2)$ as
\begin{alignat}{3}
  (x^7 f_1)' = - \frac{1 - x^7}{2} \Big\{ \frac{x^7 ( h_1 - h_2 )}{1 + \alpha^7 x^7} \Big\}' 
  + \frac{362880}{x^{28}} ( 5317 - 4254 x^7 ). \label{eq:f1'}
\end{alignat}
There remain 4 equations to be solved.

\subsection{Solve $E_2 + E_3 = 0$}

Next we evaluate a combination of $E_2 + E_3$.
\begin{alignat}{3}
  0 &= \frac{E_2 + E_3}{(1+ \alpha^7 x^7)^3} \notag
  \\
  &= 2 x^{28} \big\{ 8 (x^{7} f_1)' + x (x^{7} f_1)'' \big\}
  + 49 ( 3 - (6 + \alpha^7) x^7 - 2 \alpha^7 x^{14} ) \frac{x^{34}( h_1 - h_2 )}{( 1 + \alpha^7 x^7 )^3} \notag
  \\
  &\quad\,
  + ( 37 - (58 - 9 \alpha^7) x^7 - 30 \alpha^7 x^{14} ) \frac{x^{35}( h_1 - h_2)'}{( 1 + \alpha^7 x^7 )^2}
  + 2 ( 1 - x^7 ) \frac{x^{36}( h_1 - h_2 )''}{1 + \alpha^7 x^7} \notag
  \\[0.1cm]
  &\quad\,
  - 4354560 ( 3297 - 806 x^7 ) \notag
  \\
  &= 2 x^{28} \big\{ 8 (x^{7} f_1)' + x (x^{7} f_1)'' \big\} 
  + 3 ( 3 - 10 x^7 ) x^{28} \Big\{ \frac{x^7 ( h_1 - h_2)}{1 + \alpha^7 x^7} \Big\}'  \notag
  \\
  &\quad\,
  + 2 ( 1 - x^7 ) x^{29} \Big\{ \frac{x^7( h_1 - h_2 )}{1 + \alpha^7 x^7} \Big\}''
  - 4354560 ( 3297 - 806 x^7 ) \notag
  \\
  &= ( 1 - x^7 ) x^{29} \Big\{ \frac{x^7 ( h_1 - h_2 )}{1 + \alpha^7 x^7} \Big\}''
  + ( 1 - 15 x^7 ) x^{28} \Big\{ \frac{x^7 ( h_1 - h_2 )}{1 + \alpha^7 x^7} \Big\}' \notag
  \\[0.1cm]
  &\quad\,
  - 1451520 ( 63061 - 30069 x^7 ) \notag
  \\
  &= \frac{x^{28}}{1 - x^7} 
  \Big[ x ( 1 - x^7)^2 \Big\{ \frac{x^7 ( h_1 - h_2 )}{1 + \alpha^7 x^7} \Big\}' \Big]'
  - 1451520 ( 63061 - 30069 x^7 ).
\end{alignat}
Note that the eq.~(\ref{eq:f1'}) is employed to eliminate $f_1(x)$ out of the equation.
It is possible to integrate the above equation once, and the result becomes
\begin{alignat}{3}
  \Big\{ \frac{x^7 ( h_1 - h_2 )}{1 + \alpha^7 x^7} \Big\}' 
  &= \frac{1451520}{( 1 - x^7 )^2} \Big( - \frac{63061}{27x^{28}} 
  + \frac{9313}{2x^{21}} - \frac{2313}{x^{14}} \Big)
  + \frac{c_1}{x ( 1 - x^7 )^2} \notag
  \\
  &= 1451520 \Big\{ - \frac{63061}{27 x^{28}} - \frac{793}{54 x^{21}}
  - \frac{61}{9 x^{14}} + \frac{61}{54 x^7} 
  + \frac{61 (8 - x^7)}{54 ( 1 - x^7 )^2} \Big\} \label{eq:h1'-h2'}
  \\
  &\quad\,
  + \frac{c_1}{x ( 1 - x^7 )^2}, \notag
\end{alignat}
where $c_1$ is an integral constant.
$h_1(x) - h_2(x)$ can be derived by integrating the eq.~(\ref{eq:h1'-h2'}).
In order to execute the integral, we define the following function,
\begin{alignat}{3}
  I (x) &= \log \frac{x^7(x-1)}{x^7-1} 
  - \sum_{n=1,3,5} \cos \frac{n\pi}{7} \log \Big( x^2 + 2 x \cos \frac{n\pi}{7} + 1 \Big) \notag
  \\[-0.1cm]
  &\quad\,
  - 2 \sum_{n=1,3,5} \sin \frac{n\pi}{7} \bigg\{
  \tan^{-1} \bigg( \frac{x + \cos \tfrac{n\pi}{7} }
  {\sin \tfrac{n\pi}{7}} \bigg) - \frac{\pi}{2} \bigg\}. \label{eq:I}
\end{alignat}
It is useful to note following relations.
\begin{alignat}{3}
  &I'(x) = \frac{7(1-x)}{x(1-x^7)}, \qquad
  \Big\{ \log \Big( 1 - \frac{1}{x^7} \Big) \Big\}'= - \frac{7}{x(1-x^7)}, \notag
  \\
  &\Big\{ \frac{1}{7} I(x) + \frac{1}{7} \log \Big( 1 - \frac{1}{x^7} \Big) \Big\}' 
  = -\frac{1}{1-x^7}, \notag
  \\
  &\bigg\{ \frac{1}{7} \frac{x}{x^7-1} 
  + \frac{6}{49} I(x) + \frac{6}{49} \log \Big( 1 - \frac{1}{x^7} \Big) \bigg\}' 
  = - \frac{1}{( 1 - x^7 )^2}, \label{eq:relI}
  \\
  &\bigg\{ \frac{1}{7} \frac{1}{x^7-1} 
  + \frac{1}{7} \log \Big( 1 - \frac{1}{x^7} \Big) \bigg\}' 
  = - \frac{1}{x( 1 - x^7 )^2}. \notag
\end{alignat}
By using these relations it is possible to integrate the eq.~(\ref{eq:h1'-h2'}), 
and the result becomes
\begin{alignat}{3}
  \frac{h_1 - h_2}{1 + \alpha^7 x^7} &= 
  \frac{1130053120}{9 x^{34}} + \frac{1065792}{x^{27}}
  + \frac{9838080}{13 x^{20}} - \frac{273280}{x^{13}}
  - \frac{1639680 ( x - \tilde{c}_1 )}{x^7(x^7-1)} \notag
  \\
  &\quad\,
  - \frac{1639680}{x^7} I(x) 
  + \frac{c_2}{x^7}
  - (1 - \tilde{c}_1) \frac{1639680}{x^7} \log \Big( 1 - \frac{1}{x^7} \Big), 
\end{alignat}
where $c_2$ is an integral constant and $c_1$ is redefined as $c_1 = - 11477760 \tilde{c}_1$.
Because $h_1(x) - h_2(x)$ should be finite at $x = 1$, we choose $\tilde{c}_1 = 1$.
Then $h_1(x) - h_2(x)$ takes the form of
\begin{alignat}{3}
  h_1 - h_2 &= \frac{1130053120}{9 x^{34}}
  + \frac{448 ( 21411 + 2522440 \alpha^7 )}{9 x^{27}}
  + \frac{81984 (120 + 169 \alpha^7)}{13 x^{20}} \notag
  \\
  &\quad\,
  - \frac{273280 (13 - 36 \alpha^7) }{13x^{13}}
  - \frac{1639680}{x^7} 
  + \frac{273280 (6 - \alpha^7)}{x^6} \notag
  \\
  &\quad\,
  - \frac{1639680 (1 + \alpha^7) (x - 1)}{x^7 - 1}
  - 1639680 \Big( \frac{1}{x^7} + \alpha^7 \Big) I(x)
  + \frac{c_2}{x^7} + c_2 \alpha^7. \label{eq:h1-h2}
\end{alignat}

Now we are ready to derive an explicit form of $f_1(x)$.
By inserting the eq.~(\ref{eq:h1'-h2'}) into the eq.~(\ref{eq:f1'}), 
we obtain differential equation for $f_1(x)$,
\begin{alignat}{3}
  (x^7 f_1)' &= 725760 \bigg\{(1 - x^7) 
  \Big( \frac{63061}{27 x^{28}} + \frac{793}{54 x^{21}} + \frac{61}{9 x^{14}} - \frac{61}{54 x^7} \Big)
  - \frac{61 (8 - x^7)}{54 ( 1 - x^7 )} \bigg\} \notag
  \\
  &\quad\,
  + \frac{5738880}{x ( 1 - x^7 )}
  + \frac{362880}{x^{28}} ( 5317 - 4254 x^7 ) \notag
  \\
  &= 725760 \Big( \frac{269681}{54 x^{28}} - \frac{240187}{54 x^{21}} 
  - \frac{427}{54 x^{14}} - \frac{427}{54 x^7} \Big)
  + \frac{5738880 (1 - x)}{x(1 - x^7)} . \label{eq:f1'2}
\end{alignat}
By using the relations (\ref{eq:relI}), $f_1(x)$ can be solved as
\begin{alignat}{3}
  f_1 &= - \frac{1208170880}{9x^{34}} 
  + \frac{161405664}{x^{27}} + \frac{5738880}{13 x^{20}} + \frac{956480}{x^{13}} + \frac{c_3}{x^7}
  + \frac{819840}{x^7} I(x), \label{eq:f1}
\end{alignat}
where $c_3$ is an integral constant.
So far we solved $E_1 - E_2 + E_3 - E_4 = 0$ and $E_2 + E_3 = 0$ to obtain $h_1(x) - h_2(x)$ and $f_1(x)$.
There remain 3 equations to be solved.

\subsection{Solve $E_1 + E_2 = 0$}

A linear combination of $E_1 + E_2$ is calculated as follows.
\begin{alignat}{3}
  0 &= - \frac{E_1 + E_2}{2x^{28}(1-x^7)(1+ \alpha^7 x^7)^2} \notag
  \\
  &= - \frac{9 + 16 \alpha^7 x^7}{2} \Big\{ \frac{x^7 ( h_1 - h_2 )}{1 + \alpha^7 x^7} \Big\}'
  + \frac{34836480}{x^{28}} ( 67 - (910 + 941 \alpha^7) x^7 ) - (x^8 h_2')' \notag
  \\
  &= 
  725760 (9 + 16 \alpha^7 x^7) \bigg\{ \frac{63061}{27 x^{28}} + \frac{793}{54 x^{21}}
  + \frac{61}{9 x^{14}} - \frac{61}{54 x^7}
  - \frac{61 (8 - x^7)}{54 ( 1 - x^7 )^2} \bigg\} \notag
  \\
  &\quad\,
  + \frac{5738880 (9 + 16 \alpha^7 x^7)}{x ( 1 - x^7 )^2}
  + \frac{34836480}{x^{28}} ( 67 - (910 + 941 \alpha^7) x^7 ) - (x^8 h_2')' \notag
  \\
  &= 725760 \bigg( \frac{72709}{3 x^{28}} 
  - \frac{2351583 + 421120 \alpha^7}{54 x^{21}} 
  + \frac{1647 + 6344 \alpha^7}{27 x^{14}} 
  - \frac{183 - 1952 \alpha^7}{18 x^7} \bigg) \notag
  \\
  &\quad\,
  + 819840 (9 + 16 \alpha^7) \bigg\{ \frac{7 (1 - x)}{x ( 1 - x^7 )^2} - \frac{1}{1 - x^7} \bigg\} 
  - \frac{91822080 \alpha^7 (1-x)}{x ( 1 - x^7 )} - (x^8 h_2')' .
\end{alignat}
Here we employed the eq.~(\ref{eq:h1'-h2'}).
This is a differential equation only on $h_2(x)$, and it is possible to integrate it once
by employing the relations (\ref{eq:relI}).
The result is calculated as
\begin{alignat}{3}
  h_2' &= - \frac{651472640}{x^{35}} 
  + \frac{672 (2351583 + 421120 \alpha^7)}{x^{28}}
  - \frac{1639680 (27 + 104 \alpha^7)}{13x^{21}} \notag
  \\
  &\quad\,
  + \frac{409920 (3 - 32 \alpha^7)}{x^{14}} 
  + 819840 (9 + 16 \alpha^7) \frac{1-x}{x^8(1-x^7)} 
  + \frac{7378560}{x^8} I(x)
  - \frac{7 c_4}{x^8} \notag
  \\
  &= - \frac{651472640 }{x^{35}} + \frac{672 (2351583 + 421120 \alpha^7)}{x^{28}}
  - \frac{1639680 (27 + 104 \alpha^7)}{13x^{21}} \notag
  \\
  &\quad\,
  + \frac{409920 (3 - 32 \alpha^7)}{x^{14}} 
  + 1639680 (9 + 8 \alpha^7) \frac{1-x}{x^8(1-x^7)} 
  + \Big( - \frac{1054080}{x^7} I(x) \Big)'
  - \frac{7 c_4}{x^8} \notag
  \\
  &= - \frac{651472640}{x^{35}} + \frac{672 (2351583 + 421120 \alpha^7)}{x^{28}}
  - \frac{1639680 (27 + 104 \alpha^7)}{13x^{21}} \notag
  \\
  &\quad\,
  + \frac{409920 (3 - 32 \alpha^7)}{x^{14}} 
  + \frac{1639680 (9 + 8 \alpha^7)}{x^8}
  - \frac{1639680 (9 + 8 \alpha^7)}{x^7} \notag
  \\
  &\quad\,
  + \Big\{ 1054080 \Big( 2 - \frac{1}{x^7} + \frac{16\alpha^7}{9} \Big) I(x) \Big\}' 
  - \frac{7 c_4}{x^8},
\end{alignat}
where $c_4$ is an integral constant.
Now it is easy to integrate the above differential equation, and $h_2(x)$ is derived as
\begin{alignat}{3}
  h_2 &= \frac{19160960}{x^{34}} - \frac{224(2351583 + 421120 \alpha^7)}{9 x^{27}}
  + \frac{81984 (27 + 104 \alpha^7)}{13 x^{20}} \notag
  \\
  &\quad\,
  - \frac{409920(3 - 32 \alpha^7)}{13 x^{13}}
  - \frac{234240 (9 + 8 \alpha^7)}{x^7} + \frac{273280 (9 + 8 \alpha^7)}{x^6} \notag
  \\
  &\quad\,
  + 1054080 \Big( 2 - \frac{1}{x^7} + \frac{16\alpha^7}{9} \Big) I(x)
  + \frac{c_4}{x^7} + c_5, \label{eq:h2sol}
\end{alignat}
where $c_5$ is an integral constant.
Inserting this result into the eq.~(\ref{eq:h1-h2}), we obtain
\begin{alignat}{3}
  h_1 &= \frac{1302501760}{9x^{34}} 
  - \frac{224(2308761 - 4623760 \alpha^7) }{9x^{27}}
  + \frac{1721664 (7 + 13 \alpha^7)}{13x^{20}} \notag
  \\
  &\quad\,
  - \frac{956480 (5 - 24 \alpha^7)}{13 x^{13}}
  - \frac{1873920 (2 + \alpha^7)}{x^7}
  + \frac{273280 (15 + 7 \alpha^7)}{x^6} \notag
  \\
  &\quad\,
  - 1639680 (1 + \alpha^7) \frac{x-1}{x^7-1}
  + 117120 \Big( 18 - \frac{23}{x^7} + 2 \alpha^7 \Big) I(x) \notag
  \\
  &\quad\,
  + \frac{c_2 + c_4}{x^7} + c_2 \alpha^7 + c_5. \label{eq:h1sol}
\end{alignat}
We have already solved $E_1 - E_2 + E_3 - E_4 = 0$, $E_2 + E_3 = 0$ and $E_1 + E_2 = 0$ 
to obtain $h_1(x)$, $h_2(x)$ and $f_1(x)$.
There remain 2 equations to be solved.

\subsection{Solve $E_5 = 0$}

The equation $E_5 = 0$ should be solved to determine $h_3(x)$.
\begin{alignat}{3}
  0 &= \frac{E_5}{7 x^{28}(1 + \alpha^7 x^7)^2} \notag
  \\
  &= \bigg\{ 
  - \frac{2 x^{7} (h_2 + h_3)}{1 + \alpha^7 x^7} - \frac{x^8}{7} (h_2 + h_3)' 
  + \frac{x^7 (h_1 + 3 h_2)}{1 + \alpha^7 x^7} 
  - \frac{371589120}{x^{27}} + \frac{452874240}{x^{20}} \bigg\}' \notag
  \\
  &= \bigg[
  - \frac{(1 + \alpha^7 x^7)^2}{7x^6} 
  \bigg\{ \frac{x^{14}(h_2 + h_3)}{(1 + \alpha^7 x^7)^2} \bigg\}'
  + \frac{x^7 (h_1 + 3 h_2)}{1 + \alpha^7 x^7} 
  - \frac{371589120}{x^{27}} + \frac{452874240}{x^{20}} \bigg]'.
\end{alignat}
This can be easily integrated once, and by using the eqs.~(\ref{eq:h2sol}) and (\ref{eq:h1sol})
we obtain a differential equation for $h_2(x) + h_3(x)$.
\begin{alignat}{3}
  &\bigg\{ \frac{x^{14}(h_2 + h_3)}{(1 + \alpha^7 x^7)^2} \bigg\}' \notag
  \\
  &= \frac{7 x^{13}}{(1 + \alpha^7 x^7)^3} 
  \bigg\{ h_1 + 3h_2 - \frac{371589120(1 + \alpha^7 x^7)}{x^{34}} 
  + \frac{452874240(1 + \alpha^7 x^7)}{x^{27}} 
  + \frac{c_6 (1 + \alpha^7 x^7)}{x^7} \bigg\} \notag
  \\
  &= \frac{7 x^{13}}{(1 + \alpha^7 x^7)^3} 
  \bigg\{ - \frac{1524454400}{9 x^{34}} 
  + \frac{2240 (883233 - 1156952 \alpha^7)}{9 x^{27}}
  + \frac{4032 (4636 + 1472055 \alpha^7)}{13 x^{20}} \notag
  \\
  &\quad\,
  - \frac{273280 (31 - 228 \alpha^7)}{13 x^{13}}
  - \frac{234240 (43 + 32 \alpha^7)}{x^7} 
  + \frac{273280 (42 + 31 \alpha^7)}{x^6}
  - \frac{1639680 (1 + \alpha^7) (x - 1)}{x^7 - 1} \notag
  \\
  &\quad\,
  + 234240 \Big( 36 - \frac{25}{x^7} + 25 \alpha^7 \Big) I(x)
  + \frac{c_2 + 4 c_4 + c_6}{x^7} + c_2 \alpha^7 + 4 c_5 + c_6 \alpha^7
  \bigg\} \notag
  \\
  &= \bigg[ \frac{x^{14}}{(1 + \alpha^7 x^7)^2} 
  \bigg\{ \frac{533559040}{9 x^{34}} 
  - \frac{118368320}{x^{27}}
  - \frac{21807744}{13 x^{20}}
  + \frac{3747840}{x^{14}}
  - \frac{59301760}{13 x^{13}}
  - \frac{4216320}{x^7} \notag
  \\
  &\qquad\,
  + \frac{4919040}{x^6}
  + 234240 \Big( 18 - \frac{25}{x^7} \Big) I(x) 
  + \frac{c_2 + 4 c_4 + c_6}{x^7} 
  + c_2 \alpha^7 + 2 c_4 \alpha^7 + 2 c_5 + c_6 \alpha^7 \bigg\} \bigg]' \notag
\end{alignat}
where $c_6$ is an integral constant. In order to obtain the last equality we used relations (\ref{eq:relI}).
This is a differential equation only on $h_2(x) + h_3(x)$ and we obtain
\begin{alignat}{3}
  h_2 + h_3 &= \frac{533559040}{9 x^{34}} - \frac{118368320}{x^{27}}
  - \frac{21807744}{13 x^{20}} + \frac{3747840}{x^{14}} \notag
  \\
  &\quad\,
  - \frac{59301760}{13 x^{13}} - \frac{4216320}{x^7} + \frac{4919040}{x^6}
  + 234240 \Big( 18 - \frac{25}{x^7} \Big) I(x) \notag
  \\
  &\quad\,
  + \frac{c_2 + 4 c_4 + c_6}{x^7} 
  + c_2 \alpha^7 + 2 c_4 \alpha^7 + 2 c_5 + c_6 \alpha^7 + c_g \Big( \alpha^7 + \frac{1}{x^7} \Big)^2,
\end{alignat}
where $c_g$ is an integral constant.
As explained in eq.~(\ref{eq:gaugetr}), $c_g$ corresponds to the parameter of 
the coordinate transformation $dz \to dz + c_g dt$.
Since the explicit form of $h_2(x)$ is given by the eq.~(\ref{eq:h2sol}), $h_3(x)$ is expressed as
\begin{alignat}{3}
  h_3 &= \frac{361110400}{9 x^{34}}
  - \frac{224(2404287 - 421120 \alpha^7) }{9 x^{27}}
  - \frac{81984(293 + 104 \alpha^7) }{13 x^{20}} \notag
  \\
  &\quad\,
  + \frac{3747840}{x^{14}}
  - \frac{136640(425 + 96 \alpha^7) }{13 x^{13}}
  - \frac{234240(9 - 8 \alpha^7) }{x^7} \notag
  \\
  &\quad\,
  + \frac{273280 (9 - 8 \alpha^7) }{x^6} 
  + 117120 \Big( 18 - \frac{41}{x^7} - 16 \alpha^7 \Big) I(x) \notag
  \\
  &\quad\,
  + \frac{c_2 + 3 c_4 + c_6}{x^7}
  + c_2 \alpha^7 + 2 c_4 \alpha^7 + c_5 + c_6 \alpha^7 
  + c_g \Big( \alpha^7 + \frac{1}{x^7} \Big)^2. \label{eq:h3sol}
\end{alignat}
Now we have solved 4 equations and derived $h_i(x)$ and $f_1(x)$.
There remain only one equation to be solved.

\subsection{Solve $E_1 = 0$}

The final independent equation which should be solved is $E_1 = 0$.
By inserting solutions obtained so far, the equation $E_1 = 0$ gives
a relation among integral constants.
\begin{alignat}{3}
  - \frac{E_1}{49 \alpha^7 x^{34} (1 + x^7)} 
  = 1873920 \alpha^{14} + c_3 \alpha^7 - c_4 (2 + \alpha^7) - c_5 - c_6 (1 + \alpha^7) = 0.
\end{alignat}
From this, $c_3$ is expressed as
\begin{alignat}{3}
  c_3 \alpha^7 = - 1873920 \alpha^{14} + c_4 (2 + \alpha^7) 
  + c_5 + c_6 (1 + \alpha^7). \label{eq:relc}
\end{alignat}
There remain 4 integral constants $c_2$, $c_4$, $c_5$ and $c_6$ in $h_i(x)$, which should be fixed by
imposing boundary conditions and requiring consistencies with various limits in table \ref{table:limits}.

\subsection{Determination of integral constants}

Our remaining task is to determine integral constants $c_2$, $c_4$, $c_5$ and $c_6$ in $h_i(x)$.
First of all, the solution should be asymptotically flat.
This means that when $x$ goes to the infinity, $h_1(x)$, $h_2(x)$ and $f_1(x)$ should vanish
and $h_2(x) + h_3(x)$ should do up to the coordinate transformation (\ref{eq:gaugetr}).
Since the function $I(x)$ is expanded around $x \sim \infty$ as
\begin{alignat}{3}
  I(x) &= - \frac{7}{6 x^6} + \frac{1}{x^7} - \frac{7}{13 x^{13}} + \frac{1}{2 x^{14}} 
  + \mathcal{O}\Big(\frac{1}{x^{15}}\Big),
\end{alignat}
asymptotic behaviors of $h_1(x)$, $h_2(x)$, $h_2(x) + h_3(x)$ and $f_1(x)$ are evaluated like
\begin{alignat}{3}
  h_1 &= c_2 \alpha^7 + c_5 + \frac{c_2 + c_4}{x^7} 
  + \frac{1756800 \alpha^7}{x^{14}} + \mathcal{O}\Big(\frac{1}{x^{15}}\Big) , \notag
  \\[0.2cm]
  h_2 &= c_5 + \frac{c_4}{x^7} + \frac{936960 \alpha^7}{x^{14}}
  + \mathcal{O}\Big(\frac{1}{x^{15}}\Big), 
  \\[0.2cm]
  h_2 + h_3 &= c_2 \alpha^7 + 2 c_5 + c_6 \alpha^7 
  + \frac{c_2 + c_6}{x^7} 
  - \frac{2 c_4}{\alpha^7 x^{14}}
  + \Big(c_g + \frac{2 c_4}{\alpha^7} \Big) \Big( \alpha^7 + \frac{1}{x^7} \Big)^2 
  + \mathcal{O}\Big(\frac{1}{x^{15}}\Big), \notag
  \\[0.2cm]
  f_1 &= \frac{c_3}{x^7} + \frac{819840}{x^{14}}
  + \mathcal{O}\Big(\frac{1}{x^{15}}\Big) . \notag
\end{alignat}
Due to the asymptotic flatness, constants in $h_1(x)$ and $h_2(x)$ should vanish in the above equations.
Therefore we obtain
\begin{alignat}{3}
  c_2 = c_5 = 0,
\end{alignat}
and asymptotic behaviors of $h_1(x)$, $h_2(x)$, $h_2(x) + h_3(x)$ and $f_1(x)$ become
\begin{alignat}{3}
  h_1 &= \frac{c_4}{x^7} 
  + \frac{1756800 \alpha^7}{x^{14}} + \mathcal{O}\Big(\frac{1}{x^{15}}\Big) , \notag
  \\[0.2cm]
  h_2 &= \frac{c_4}{x^7} + \frac{936960 \alpha^7}{x^{14}}
  + \mathcal{O}\Big(\frac{1}{x^{15}}\Big), 
  \\[0.2cm]
  h_2 + h_3 &= 
  - \frac{c_6}{x^7} 
  - \frac{2 c_4 + c_6}{\alpha^7 x^{14}}
  + \Big( c_g + \frac{2 c_4 + c_6}{\alpha^7} \Big) \Big( \alpha^7 + \frac{1}{x^7} \Big)^2 
  + \mathcal{O}\Big(\frac{1}{x^{15}}\Big), \notag
  \\[0.2cm]
  f_1 &= \frac{c_3}{x^7} + \frac{819840}{x^{14}}
  + \mathcal{O}\Big(\frac{1}{x^{15}}\Big) . \notag
\end{alignat}
Note that $f_1(x)$ automatically goes to zero when $x \to \infty$.

Below we consider consistency conditions with the extremal limit and the Schwarzschild one 
in the table \ref{table:limits}.
In the extremal limit $\alpha \to 0$, the classical geometry preserves a half supersymmetry
and does not receive any higher curvature corrections~\cite{KR}.
Therefore the mass and the charge are fixed to be the same, and we require
$\frac{1}{\alpha^6} f_1(\frac{r}{r_-\alpha})$ should be zero and
$\frac{1}{\alpha^{13}}(h_2(\frac{r}{r_-\alpha}) + h_3(\frac{r}{r_-\alpha}))$ 
should vanish up to the gauge transformation.
These conditions are satisfied if $c_3 \sim \alpha^n (n \geq 0)$ and $c_4 \sim c_6 \sim \alpha^n (n \geq 7)$.

On the other hand, by taking the Schwarzschild limit $r_- \to 0$ with $r_- \alpha$ fixed,
$f_1(\frac{r}{r_-\alpha})$ should be finite and
$r_-^7(h_2(\frac{r}{r_-\alpha}) + h_3(\frac{r}{r_-\alpha}))$ should vanish up to the gauge transformation.
These conditions are satisfied if $c_3 \sim \alpha^n (n\leq 0)$, $c_4 \sim \alpha^n (n \leq 13)$ 
and $c_6 \sim \alpha^n (n \leq 6)$.
Combining these restrictions with the eq.~(\ref{eq:relc}), we obtain
\begin{alignat}{3}
  c_3 = 3747840, \qquad c_4 = 1873920 \alpha^7, \qquad c_6 = 0,
\end{alignat}
and finally asymptotic behaviors of $h_1(x)$, $h_2(x)$, $h_2(x) + h_3(x)$ and $f_1(x)$ become
\begin{alignat}{3}
  h_1 &= \frac{1873920 \alpha^7}{x^7} 
  + \frac{1756800 \alpha^7}{x^{14}} + \mathcal{O}\Big(\frac{1}{x^{15}}\Big) , \notag
  \\[0.2cm]
  h_2 &= \frac{1873920 \alpha^7}{x^7} + \frac{936960 \alpha^7}{x^{14}}
  + \mathcal{O}\Big(\frac{1}{x^{15}}\Big), \label{eq:asym}
  \\[0.2cm]
  h_2 + h_3 &= 
  - \frac{3747840}{x^{14}}
  + \big( c_g + 3747840 \big) \Big( \alpha^7 + \frac{1}{x^7} \Big)^2 
  + \mathcal{O}\Big(\frac{1}{x^{15}}\Big), \notag
  \\[0.2cm]
  f_1 &= \frac{3747840}{x^7} + \frac{819840}{x^{14}}
  + \mathcal{O}\Big(\frac{1}{x^{15}}\Big) . \notag
\end{alignat}
Therefore all integral constants except $c_g$ are uniquely determined by imposing asymptotic flatness and
requiring consistencies with the extremal limit and the Schwarzschild one.

\section{3 Limits of the Quantum Black 0-brane}\label{sec:sol}

\subsection{The quantum black 0-brane solution}

Let us summarize the quantum black 0-brane solution.
The dimensional reduction of the quantum M-wave solution is identified with
the quantum black 0-brane in 10 dimensions, and the geometry is described by
\begin{alignat}{3}
  &ds_{10}^2 = - H_1^{-1} H_2^{\frac{1}{2}} F_1 dt^2 + H_2^{\frac{1}{2}} F_1^{-1} dr^2
  + H_2^{\frac{1}{2}} r^2 d\Omega_8^2, \label{eq:QB0}
  \\
  &e^\phi = H_2^{\frac{3}{4}}, \quad
  C = \sqrt{1+\alpha^7} (H_2 H_3)^{-\frac{1}{2}} dt, \notag
\end{alignat}
where
\begin{alignat}{3}
  &H_i = 1 + \frac{r_-^7}{r^7} + \frac{\gamma}{r_-^6 \alpha^{13}} h_i \Big(\frac{r}{r_-\alpha}\Big), \qquad
  F_1 = 1 - \frac{r_-^7 \alpha^7}{r^7} + \frac{\gamma}{r_-^6 \alpha^6} f_1 \Big(\frac{r}{r_-\alpha}\Big). 
  \label{eq:solg1}
\end{alignat}
And the functions $h_i(x)$ and $f_1(x)$ are uniquely determined as
\begin{alignat}{3}
  h_1(x) &= \frac{1302501760}{9x^{34}} 
  - \frac{224(2308761 - 4623760 \alpha^7) }{9x^{27}}
  + \frac{1721664(7 + 13 \alpha^7)}{13x^{20}} \notag
  \\
  &\quad\,
  - \frac{956480(5 - 24 \alpha^7)}{13 x^{13}}
  - \frac{3747840}{x^7}
  + \frac{273280(15 + 7 \alpha^7)}{x^6} \notag
  \\
  &\quad\,
  - 1639680 (1 + \alpha^7) \frac{x-1}{x^7-1}
  + 117120 \Big( 18 - \frac{23}{x^7} + 2 \alpha^7 \Big) I(x), \notag
  \\[0.2cm]
  h_2(x) &= \frac{19160960}{x^{34}} - \frac{224(2351583 + 421120 \alpha^7)}{9 x^{27}}
  + \frac{81984(27 + 104 \alpha^7)}{13 x^{20}} \notag
  \\
  &\quad\,
  - \frac{409920(3 - 32 \alpha^7)}{13 x^{13}}
  - \frac{2108160}{x^7} 
  + \frac{273280(9 + 8 \alpha^7)}{x^6} \notag
  \\
  &\quad\,
  + 1054080 \Big( 2 - \frac{1}{x^7} + \frac{16}{9} \alpha^7 \Big) I(x), \label{eq:solg2}
  \\[0.2cm]
  h_3(x) &= \frac{361110400}{9 x^{34}}
  - \frac{224(2404287 - 421120 \alpha^7) }{9 x^{27}}
  - \frac{81984(293 + 104 \alpha^7) }{13 x^{20}} \notag
  \\
  &\quad\,
  - \frac{136640(425 + 96 \alpha^7) }{13 x^{13}}
  - \frac{2108160}{x^7}
  + \frac{273280(9 - 8 \alpha^7) }{x^6} \notag
  \\
  &\quad\,
  + 117120 \Big( 18 - \frac{41}{x^7} - 16 \alpha^7 \Big) I(x), \notag
  \\[0.2cm]
  f_1(x) &= - \frac{1208170880}{9x^{34}} 
  + \frac{161405664}{x^{27}} 
  + \frac{5738880}{13 x^{20}}
  + \frac{956480}{x^{13}}
  + \frac{3747840}{x^7} 
  + \frac{819840}{x^7} I(x), \notag
\end{alignat}
where the function $I(x)$ is defined by the eq.~(\ref{eq:I}).
Note that $h_2(x) + h_3(x)$ and $f_1(x)$ do not depend on $\alpha$.
Below we examine 3 limits in table \ref{table:limits}.

\subsection{The extremal limit}

In the extremal case, the classical solution is described by the eq.~(\ref{eq:11sol}) with $r_+=r_-$.
This is a half BPS solution and does not receive any higher curvature corrections~\cite{KR}.
This is explicitly checked by taking the extremal limit $\alpha \to 0$ to the solution (\ref{eq:QB0})
with (\ref{eq:solg1}) and (\ref{eq:solg2}). 
In fact, because of the asymptotic behaviors (\ref{eq:asym}), all quantum corrections are irrelevant
and the solution is given by the eq.~(\ref{eq:QB0}) with
\begin{alignat}{3}
  H_i = 1 + \frac{r_-^7}{r^7}, \qquad F_1 = 1.
\end{alignat}
Thus the extremal solution coincides with a half BPS solution and
is not affected by the leading quantum correction at 1-loop.

\subsection{The Schwarzschild limit}

The Schwarzschild limit is given by $r_- \to 0$ with $r_- \alpha$ fixed.
The solution is given by the eq.~(\ref{eq:QB0}) with
\begin{alignat}{3}
  H_i &= 1 + \frac{\gamma}{r_-^6 \alpha^6} \hat{h}_i \Big(\frac{r}{r_-\alpha}\Big), \qquad
  F_1 = 1 - \frac{r_-^7 \alpha^7}{r^7} + \frac{\gamma}{r_-^6 \alpha^6} f_1 \Big(\frac{r}{r_-\alpha}\Big), 
  \label{eq:solS1}
\end{alignat}
and
\begin{alignat}{3}
  \hat{h}_1(x) &= \frac{1035722240}{9x^{27}} \!+\! \frac{1721664}{x^{20}} 
  \!+\! \frac{22955520}{13 x^{13}} \!+\! \frac{1912960}{x^6} 
  \!-\! 1639680 \frac{x-1}{x^7-1} 
  \!+\! 234240 I(x), \notag
  \\[0.2cm]
  \hat{h}_2(x) &= - \hat{h}_3(x) = 
  -\! \frac{94330880}{9 x^{27}} \!+\! \frac{655872}{x^{20}} 
  \!+\! \frac{13117440}{13 x^{13}} \!+\! \frac{2186240}{x^6} 
  \!+\! 1873920 I(x). \label{eq:solS2}
\end{alignat}
Notice that $\hat{h}_2(x) + \hat{h}_3(x) = 0$, which means that R-R 1-form gauge field 
is trivial and the R-R charge is zero.

\subsection{The near horizon limit}

If we take the near horizon limit $\alpha \to 0$ with $\frac{r}{r_-\alpha}$ and 
$\frac{\gamma}{r_-^6\alpha^6}$ fixed, 
terms of higher powers of $\alpha$ in the eq.~(\ref{eq:solg2}) vanish and the solution is
given by the eq.~(\ref{eq:QB0}) with
\begin{alignat}{3}
  H_i &= \frac{r_-^7}{r^7} + \frac{\gamma}{r_-^6 \alpha^{13}} \tilde{h}_i \Big(\frac{r}{r_-\alpha}\Big), \qquad
  F_1 = 1 - \frac{r_-^7 \alpha^7}{r^7} + \frac{\gamma}{r_-^6 \alpha^6} f_1 \Big(\frac{r}{r_-\alpha}\Big), 
  \label{eq:soln1}
\end{alignat}
and
\begin{alignat}{3}
  \tilde{h}_1(x) &= \frac{1302501760}{9x^{34}} 
  - \frac{57462496}{x^{27}}
  + \frac{12051648}{13x^{20}}
  - \frac{4782400}{13 x^{13}}
  - \frac{3747840}{x^7}
  + \frac{4099200}{x^6}  \notag
  \\
  &\quad\,
  - 1639680 \frac{x-1}{x^7-1}
  + 117120 \Big( 18 - \frac{23}{x^7} \Big) I(x), \notag
  \\[0.2cm]
  \tilde{h}_2(x) &= \frac{19160960}{x^{34}} 
  - \frac{58528288}{x^{27}}
  + \frac{2213568}{13 x^{20}}
  - \frac{1229760}{13 x^{13}}
  - \frac{2108160}{x^7} 
  + \frac{2459520}{x^6} \notag
  \\
  &\quad\,
  + 1054080 \Big( 2 - \frac{1}{x^7} \Big) I(x), \label{eq:soln2}
  \\[0.2cm]
  \tilde{h}_3(x) &= \frac{361110400}{9 x^{34}}
  - \frac{59840032}{x^{27}}
  - \frac{24021312}{13 x^{20}}
  - \frac{58072000}{13 x^{13}}
  - \frac{2108160}{x^7}
  + \frac{2459520}{x^6} \notag
  \\
  &\quad\, 
  + 117120 \Big( 18 - \frac{41}{x^7} \Big) I(x). \notag
\end{alignat}
This is the quantum near horizon geometry of the black 0-brane which is first derived in ref.~\cite{H2}.
Notice that compared with ref.~\cite{H2} there is an additional $\frac{3747840}{x^7}$ 
term in $f_1(x)$. As is clear form the discussions so far, this term is necessary to be consistent
with the extremal limit and the Schwarzschild one, which was not obvious in ref.~\cite{H2}.
Anyway, this additional term does not affect the entropy of the black 0-brane
since it depends on $f_1(x)$ through $7f_1(1) + f_1'(1)$.

\section{Mass, R-R Charge and Internal Energy of the Quantum Black 0-brane}

In this section we evaluate the mass, the R-R charge and the internal energy of the quantum black 0-brane.
Formulae for the mass and the R-R charge including higher derivative corrections are discussed in
the appendix. As a result, we can employ usual ADM mass and charge formulae.

\subsection{The Mass}

The mass of the quantum black 0-brane is calculated by using ADM mass formula.
In order to use the formula the metric should be written in Einstein frame.
The metric in the Einstein frame $g^E_{\mu\nu}$ is written as $g^E_{\mu\nu} = e^{-\frac{\phi}{2}} g_{\mu\nu}$,
so the line element in the Einstein frame is given by
\begin{alignat}{3}
  ds_{E}^2 &= - H_1^{-1} H_2^{\frac{1}{8}} F_1 dt^2 + H_2^{\frac{1}{8}} F_1^{-1} dr^2
  + H_2^{\frac{1}{8}} r^2 d\Omega_8^2 \notag
  \\
  &= - H_1^{-1} H_2^{\frac{1}{8}} F_1 dt^2 
  + H_2^{\frac{1}{8}} \Big\{ \delta_{ij} + (F_1^{-1} - 1) \frac{x_ix_j}{r^2} \Big\} dx^i dx^j,
\end{alignat}
where $r^2=(x^1)^2 + \cdots + (x^9)^2$ and $i,j=1,\cdots,9$.
Then, by using the eq.~(\ref{eq:asym}), the deviation from the flat space-time is given by
\begin{alignat}{3}
  h_{ij} &= H_2^{\frac{1}{8}} \Big\{ \delta_{ij} 
  + (F_1^{-1} - 1) \frac{x_ix_j}{r^2} \Big\} - \delta_{ij} \notag
  \\
  &= \frac{r_-^7 + 1873920 \gamma \, r_- \alpha}{8r^7} \delta_{ij}
  + \frac{r_-^7 \alpha^7 - 3747840 \gamma \, r_- \alpha}{r^7} \frac{x_ix_j}{r^2} 
  + \mathcal{O}\Big(\frac{1}{r^{14}}\Big).
\end{alignat}
Now we are ready to apply the ADM mass formula. 
The mass of the quantum black 0-brane is evaluated as
\begin{alignat}{3}
  M &= \frac{1}{2\kappa_{10}^2} \int_{r=\infty} d\Omega_8 r^8
  (\partial_j h^{ij} - \partial^i h^j{}_j) \frac{x_i}{r} \notag
  \\
  &= \frac{V_{S^8}}{2\kappa_{10}^2} \big\{ 8 r_-^7 \alpha^7 + 7 r_-^7 
  - 16865280 \gamma \, r_- \alpha \big\}. \label{eq:mass}
\end{alignat}
Thus the mass receives the nontrivial quantum correction at 1-loop level.

\subsection{The R-R charge}

The black 0-brane couples to the R-R 1-form field and carries the R-R charge.
From the eq.~(\ref{eq:QB0}), the R-R 1-form field $C$ and 2-form field strength $G_2 = dC$ are given by
\begin{alignat}{3}
  C &= \Big(\frac{r_+}{r_-}\Big)^\frac{7}{2} H_2^{-\frac{1}{2}} H_3^{-\frac{1}{2}} dt, \notag
  \\
  G_2 &= \frac{1}{2} \Big(\frac{r_+}{r_-}\Big)^\frac{7}{2} H_2^{-\frac{1}{2}} H_3^{-\frac{1}{2}}
  \Big( H_2^{-1} \frac{d H_2}{dr} + H_3^{-1} \frac{d H_3}{dr} \Big) dt \wedge dr.
\end{alignat}
Then the R-R charge of the quantum black 0-brane is evaluated as
\begin{alignat}{3}
  Q &= \frac{1}{2\kappa_{10}^2} \int_{r=\infty} \ast G_2 \notag
  \\
  &= \frac{1}{2\kappa_{10}^2} \int_{r=\infty} d\Omega_8 \Big\{ - \frac{1}{2} 
  \Big(\frac{r_+}{r_-}\Big)^\frac{7}{2} H_1^{\frac{1}{2}} H_2 H_3^{-\frac{1}{2}}
  \Big( H_2^{-1} \frac{d H_2}{dr} + H_3^{-1} \frac{d H_3}{dr} \Big) r^8 \Big\} \notag
  \\
  &= \frac{1}{2\kappa_{10}^2} \int_{r=\infty} d\Omega_8 
  \Big(\frac{r_+}{r_-}\Big)^\frac{7}{2} \Big[ - r^8 \frac{dH}{dr} 
  - \frac{\gamma\, r^8}{2r_-^6 \alpha^{13}} \frac{d}{dr}
  \Big\{ h_2\Big(\frac{r}{r_-\alpha}\Big) + h_3\Big(\frac{r}{r_-\alpha}\Big) \Big\} \Big] \notag
  \\
  &= \frac{V_{S^8}}{2\kappa_{10}^2} 7 r_-^7 \sqrt{1+\alpha^7}. \label{eq:charge}
\end{alignat}
Quantum corrections do not contribute to the R-R charge because 
$h_2(\frac{r}{r_-\alpha}) + h_3(\frac{r}{r_-\alpha}) \sim \frac{1}{r^{14}}$
when $r \to \infty$.
Therefore the R-R charge remains the same as the classical one.

\section{Validity of the Quantum Black 0-brane Solution}

So far we constructed the quantum black 0-brane solution by considering the effective action (\ref{eq:R4}).
After the dimensional reduction, this corresponds to the leading quantum correction 
to the type IIA supergravity.
However, this is a part of the effective action in the type IIA superstring theory,
so the black 0-brane is affected by other higher derivative corrections in general.
In this section, we examine those corrections and clarify the validity of 
the solution (\ref{eq:QB0}).

Since we compactify the 11 dimensional direction on the circle, when the radius of the circle
is finite, the Kaluza-Klein modes give nontrivial contributions to the effective action.
This corresponds to the leading tree level effective action, which is expressed by
by $e^{-2\phi} R^4$~\cite{GGV}.
It is also known that there are terms of $e^{2\phi} \partial^4 R^4$ at two loop level~\cite{GKV}.
Although the full structure of the effective action of the type IIA superstring theory
is not completed yet, from the string duality arguments, it takes the following form~\cite{GRV}
\begin{alignat}{3}
  g_s^2 e^{2\phi} \mathcal{L}
  &\sim R + \big( \alpha'^3 R^4 + \alpha'^5 \partial^4 R^4 + \cdots \big)
  + g_s^2 e^{2\phi} \big( \alpha'^3 R^4 + \alpha'^6 \partial^6 R^4 + \cdots \big) \notag
  \\
  &\qquad\,
  + (g_s^2 e^{2\phi})^2 \big( \alpha'^5 \partial^4 R^4 + \cdots \big) + \cdots 
  + (g_s^2 e^{2\phi})^n \big( \alpha'^{3+n} \partial^{2n} R^4 + \cdots \big) + \cdots, \label{eq:actexp}
\end{alignat}
where $\alpha' = \ell_s^2$.
The Riemann tensor behaves like $R_{abcd} \sim E^2$, where $E$ is a typical energy of the black 0-brane,
such as the internal energy.
Then the dimensional analysis shows that
\begin{alignat}{3}
  g_s^2 e^{2\phi} \mathcal{L}
  &\sim E^2 \Big\{ 1 + \big( \alpha'^3 E^6 + \alpha'^5 E^{10} + \cdots \big)
  + g_s^2 \big( \alpha'^3 E^6 + \alpha'^6 E^{12} + \cdots \big) \notag
  \\
  &\qquad\quad
  + g_s^4 \big( \alpha'^5 E^{10} + \cdots \big) + \cdots 
  + g_s^{2n} \big( \alpha'^{3+n} E^{6+2n} + \cdots \big) + \cdots. \label{eq:actexp2}
\end{alignat}
The leading 1-loop quantum correction $g_s^2 \alpha'^3 E^6$ becomes sub-dominant when
\begin{alignat}{3}
  1 < g_s^2 < \frac{1}{\alpha' E^2}. \label{eq:validr}
\end{alignat}
The solution (\ref{eq:QB0}) is valid in this region.
This means that the radius of the 11th direction $g_s \ell_s$ is large compared to the string scale,
and the typical energy $E$ is small compared to the Kaluza-Klein mass $\frac{1}{g_s\ell_s}$.

\section{Conclusion and Discussion}

In this paper we investigated the quantum corrections to the non-extremal M-wave and 
black 0-brane solutions. 
The effective action for the M-theory is taken into account including higher curvature $R^4$ terms, and
the equations of motion for the non-extremal M-wave are analytically solved up to the leading quantum level.
The Kaluza-Klein reduction of the quantum M-wave gives the quantum black 0-brane solution
which is asymptotically flat.
The explicit form of the solution is given by eqs.~(\ref{eq:QB0})-(\ref{eq:solg2}).
The integral constants of the solution are uniquely fixed by imposing asymptotic flatness and finiteness
around the event horizon.
We also required consistencies with the extremal limit and the Schwarzschild one 
in the table~\ref{table:limits}.
The extremal limit of the solution remains the same as the classical one, so it is not affected by
the leading quantum corrections.
The Schwarzschild limit of the quantum black 0-brane is given by eqs.~(\ref{eq:solS1}) and (\ref{eq:solS2}).
The near horizon limit is done by eqs.~(\ref{eq:soln1}) and (\ref{eq:soln2}).
Notice that the near horizon limit of the metric is the same as that in ref.~\cite{H2}
except $\frac{1}{x^7}$ term in $f_1(x)$. 
This extra term was dropped in ref.~\cite{H2} by imposing stronger boundary condition.
Nevertheless, this does not affect physical quantities, such as the near horizon limit of the internal energy.

The quantum corrections to the ADM mass and the R-R charge are discussed in the appendix.
We employed Noether and Wald's method with the vielbein formalism.
After some calculations we concluded that the corrections are suppressed at the spacial infinity and
the ADM mass and the charge formulae are still valid for the quantum black 0-brane.
The mass of the quantum black 0-brane receives the quantum correction, and the R-R charge remains the same
as the classical one. This would be a renormalization of the mass of the gravitational objects.

One of the important future works is to generalize the above discussions to the other black $p$-branes.
Although our knowledge of the effective action for the superstring theory is limited, it is possible to
consider quantum corrections to the black 6-brane, since this is uplifted to a Kaluza-Klein monopole
solution in 11 dimensions. It is also possible to construct Schwarzschild black hole solutions in various
dimensions by compactifications. 
Schwarzschild black hole in 11 dimensions is important from the viewpoint of the test of 
gauge gravity duality. The solution in 4 dimensions will also be important because it could give some 
astrophysical predictions to be observed in the future.
For other black $p$-branes, first we need to know the effective action including R-R gauge fields.

\section*{Acknowledgement}

The author would like to thank Masanori Hanada, Goro Ishiki, Hikaru Kawai and Jun Nishimura
for the motivation of this work. He also got some insights from many colleagues through discussions
on ref.~\cite{HHIN}.
This work was partially supported by the Ministry of Education, Science, 
Sports and Culture, Grant-in-Aid for Young Scientists (B) 24740140, 2012.

\appendix
\section{Higher Derivative Corrections to the ADM Mass and the R-R Charge}

Since the effective action of the M-theory (\ref{eq:R4}) contains higher curvature terms, 
the ADM mass formula and the charge of the non-extremal M-wave or the black 0-brane would 
be affected by these terms.
In this section we derive higher curvature corrections to the ADM mass and the charge via 
Noether and Wald's method \cite{Wa,IW}. We use the vielbein formalism which is important for
the supergravity\cite{H3}.

First, the variation of the Lagrangian (\ref{eq:R4}) is evaluated as
\begin{alignat}{3}
  2 \kappa_{11}^2 \dl \mathcal{L}
  &= e \big( \eta^{ac}\eta^{bd} + \gamma X^{abcd} \big) \delta R_{abcd}
  - e e^a{}_M \Big\{ R + \gamma
  \Big( t_8t_8 R^4 - \frac{1}{4!} \epsilon_{11} \epsilon_{11} R^4 \Big) \Big\} \dl e^M{}_a \notag
  \\
  &= 2 e \Big[ R_{ij} - \frac{1}{2} \eta_{ij} R + \gamma \Big\{ 
  R_{abci} X^{abc}{}_j - \frac{1}{2} \eta_{ij} 
  \Big( t_8t_8 R^4 - \frac{1}{4!} \epsilon_{11} \epsilon_{11} R^4 \Big) \Big\} \Big] \delta e^{ij} \notag
  \\
  &\quad\,
  - 2 e \big( \eta^{ac}\eta^{bd} + \gamma X^{abcd} \big) D_d \delta \omega_{cab} \notag
  \\[0.1cm]
  &= 2 e \Big[ R_{ij} - \frac{1}{2} \eta_{ij} R + \gamma \Big\{ 
  R_{abci} X^{abc}{}_j - \frac{1}{2} \eta_{ij} 
  \Big( t_8t_8 R^4 - \frac{1}{4!} \epsilon_{11} \epsilon_{11} R^4 \Big) \Big\} \Big] \delta e^{ij} \notag
  \\
  &\quad\,
  + 2 \gamma e D_d X^{abcd} \big( \delta^k_{a}\eta_{bi}\eta_{cj} 
  + \delta^k_{a} \eta_{bj} \eta_{ci} + \delta^k_c \eta_{ia} \eta_{bj} \big) D_k \delta e^{ij} \notag
  \\
  &\quad\,
  - 2 \partial_M \big\{ e \big( e^{N a} e^{M b} + \gamma X^{abNM} \big) 
  \delta \omega_{N ab} \big\} \notag
  \\[0.1cm]
  &= 2 e E_{ij} \delta e^{ij} + + \partial_M \big(e \Theta^M(\dl) \big), \label{eq:vL0}
\end{alignat}
where $\delta e^{ij} = e^i{}_M \dl e^{M j}$ and $\Theta^M(\dl)$ is defined by
\begin{alignat}{3}
  \Theta^M(\dl) &= 2 e^{M a} e^{N b} \dl \om_{N ab} 
  + 2 \gamma \big\{ X^{abMN} \dl \om_{N ab} 
  + e^{M a} ( D^b X_{aijb} + D^b X_{ajib} + D^b X_{abij} ) \delta e^{ij} \big\}. \label{eq:current1}
\end{alignat}
$E_{ij}=0$ is the equations of motion given by the eq.~(\ref{eq:MEOM}). 
The variation of the Lagrangian becomes total derivative term up to $E_{ij}=0$.
Since these variations are not covariant under the local Lorentz transformation, 
we consider following field dependent local Lorentz transformation simultaneously.
\begin{alignat}{3}
  \dl_L e^M{}_a &= - e^M{}_b \epsilon^b{}_a, \notag
  \\
  \dl_L \omega_N{}^{ab} &= - D_N \epsilon^{ab}, \label{eq:Ltr}
  \\
  \epsilon^{ab} &= \tfrac{1}{2} e^a{}_P \delta e^{P b}
  - \tfrac{1}{2} e^b{}_P \delta e^{P a}. \notag
\end{alignat}
Then the variations of fields with (\ref{eq:Ltr}) becomes
\begin{alignat}{3}
  \bar{\dl} e^M{}_a &= - \tfrac{1}{2} g^{MN} \delta g_{NP} e^P{}_a, \notag
  \\
  \bar{\dl} \om_N{}^{ab} &= \tfrac{1}{2} e^{aP} e^{bQ} 
  ( \nabla_Q \delta g_{PN} - \nabla_P \delta g_{QN} ), \label{eq:bardel}
\end{alignat}
where $\nabla_P$ represents a covariant derivative with respect to space-time indices.
These variations are covariant under local Lorentz transformation.
In the following we write $\bar{\delta}$ as $\delta$ for simplicity.
Now we consider the variation for the general coordinate transformation $x'^M = x^M - \xi^M$.
In this case the variation (\ref{eq:bardel}) becomes
\begin{alignat}{3}
  \dl_\xi e^M{}_a &= - \tfrac{1}{2} e^{M b} (D_a \xi_b + D_b \xi_a), \notag
  \\
  \dl_\xi \om_N{}^{ab} &= \tfrac{1}{2} e_{cN} \big\{ D^b(D^c \xi^a + D^a \xi^c) 
  - D^a(D^c \xi^b + D^b \xi^c) \big\} \label{eq:gctr2}
  \\
  &= \xi^P R^{ab}{}_{PN} - \tfrac{1}{2} D_N (D^a \xi^b - D^b \xi^a). \notag
\end{alignat}
Inserting the eq.~(\ref{eq:gctr2}) into the eq.~(\ref{eq:current1}), 
$\Theta^M(\xi) = \Theta^M(\dl_\xi)$ is evaluated as
\begin{alignat}{3}
  e \Theta^M(\xi) &= 2 e R^M{}_N \xi^N 
  - e e^M{}_a e^N{}_b D_N (D^a \xi^b - D^b \xi^a) \notag
  \\
  &\quad\,
  + 2 \gamma \big\{ e R_{abci} X^{abcM} \xi^i - e e^M{}_j X^{ajib} D_a D_b \xi_i
  + e e^M{}_j ( D_b X^{ijab} + D_b X^{ajib} ) D_a \xi_i \big\} \notag
  \\
  &= 2 e R^M{}_N \xi^N 
  + 2 \gamma \big\{ e R_{abci} X^{abcM} \xi^i
  - e e^M{}_j ( D_a D_b X^{ijab} + 2 D_{(a} D_{b)} X^{aijb} ) \xi_i \big\} \notag
  \\
  &\quad\,
  + \partial_N \big[ - 2 e \nabla^{[M} \xi^{N]}
  - 2 \gamma e \big\{ X^{MNPQ} \nabla_P \xi_Q
  + e^M{}_i e^N{}_a ( D_b X^{ijab} - 2 D_b X^{a(ij)b} ) \xi_j \big\} \big] \notag
  \\
  &= \partial_N \big[ - 2 e \nabla^{[M} \xi^{N]}
  - 2 \gamma e \big\{ X^{MNPQ} \nabla_P \xi_Q
  + e^M{}_i e^N{}_a ( D_b X^{ijab} - 2 D_b X^{a(ij)b} ) \xi_j \big\} \big] \notag
  \\
  &\quad\,
  + 2 \kappa_{11}^2 \xi^M \mathcal{L} + 2 e e^{M j} \xi^i E_{ij}. \label{eq:vareom}
\end{alignat}
Therefore the variation of the Lagrangian is given by
\begin{alignat}{3}
  2 \kappa_{11}^2 \delta_\xi \mathcal{L} &= \partial_M (e \Theta^M(\xi))
  + 2 e E_{ij} \delta e^{ij}, \label{eq:vL1}
  \\
  e \Theta^M(\xi) &= \pa_N \big( e Q^{MN}(\xi) \big) + 2 \kappa_{11}^2 \xi^M \mathcal{L}
  + 2 e e^{M j} \xi^i E_{ij}, \notag
  \\
  e Q^{MN}(\xi) &= - 2 e \nabla^{[M} \xi^{N]}
  - 2 \gamma e \big\{ X^{MNPQ} \nabla_P \xi_Q
  + e^M{}_i e^N{}_a ( D_b X^{ijab} - 2 D_b X^{a(ij)b} ) \xi_j \big\}. \notag
\end{alignat}

Next, let us consider the variation of the Lagrangian by using the general covariance.
Since $e^{-1}\mathcal{L}$ transforms as a scalar field, the Lagrangian does like
\begin{alignat}{3}
  \dl_\xi \mathcal{L} = \partial_M \big( \xi^M \mathcal{L} \big). \label{eq:varsym}
\end{alignat}
From the eqs.~(\ref{eq:vL1}) and (\ref{eq:varsym}), the Noether current is constructed as
\begin{alignat}{3}
  2 \kappa_{11}^2 e J^M(\xi) 
  &= e \Theta^M (\xi) - 2 \kappa_{11}^2 \xi^M \mathcal{L} 
  + \pa_N \big( e \tilde{Q}^{MN}(\xi) \big) \notag
  \\
  &= \pa_N \big\{ e \big( Q^{MN}(\xi) + \tilde{Q}^{MN}(\xi) \big) \big\}
  + 2 e e^{M j} \xi^i E_{ij}. \label{eq:curr}
\end{alignat}
Here $\tilde{Q}^{MN}(\xi)$ is an antisymmetric tensor and represents the ambiguity of the current.
In order to fix $\tilde{Q}^{MN}(\xi)$, let us consider the variation of the current.
\begin{alignat}{3}
  \dl \big( 2 \kappa_{11}^2 e J^M(\xi) \big) 
  &= \dl \big( e \Theta^M(\xi) \big) - 2 \kappa_{11}^2 \xi^M \dl \mathcal{L}
  + \pa_N \big\{ \dl \big( e \tilde{Q}^{MN}(\xi) \big) \big\} \notag
  \\
  &= \dl \big( e \Theta^M(\xi) \big) 
  - \xi^M \partial_N \big( e \Theta^N(\dl) \big) 
  - 2 e \xi^M E_{ij} \delta e^{ij} 
  + \pa_N \big\{ \dl \big( e \tilde{Q}^{MN}(\xi) \big) \big\} \notag
  \\
  &= e \om^M(\xi,\dl) + \partial_N \big\{ - 2 e \xi^{[M} \Theta^{N]}(\dl) 
  + \dl \big( e \tilde{Q}^{MN}(\xi) \big) \big\} - 2 e \xi^M E_{ij} \delta e^{ij}, \label{eq:varJ}
\end{alignat}
where we defined the symplectic current
\begin{alignat}{3}
  e \om^M(\xi,\dl) = \dl \big( e \Theta^M(\xi) \big) 
  - \dl^\text{L}_{\xi} \big( e \Theta^M(\dl) \big).
\end{alignat}
The integral of the symplectic current over Cauchy surface gives the variation of the Hamiltonian.
Therefore in order to obtain correct equations of motion, we should eliminate 
the surface term in the eq.~(\ref{eq:varJ}).
So we choose the variation of $\tilde{Q}^{MN}(\xi)$ as
\begin{alignat}{3}
  \dl \big( e \tilde{Q}^{MN}(\xi) \big) = 2 e \xi^{[M} \Theta^{N]}(\dl).
\end{alignat}
From the eq.~(\ref{eq:curr}), the variation of the current is given by
\begin{alignat}{3}
  \dl \big( 2 \kappa_{11}^2 e J^M(\xi) \big) 
  &= \pa_N \big\{ \dl \big(e Q^{MN}(\xi) \big) + 2 e \xi^{[M} \Theta^{N]}(\dl) \big\} 
  + \dl \big( 2 e e^{M j} \xi^i E_{ij} \big). \label{eq:varJ2}
\end{alignat}
Conserved quantities are obtained by integrating $e j^t(\xi)$ over 10 dimensional space.

When $\xi$ is chosen as an asymptotic time translation $\xi_T$ and
the variation of the fields, such as $\delta e^M{}_a,$ satisfy linearized equations of motion,
the variation of the mass is given by
\begin{alignat}{3}
  \delta M &= \frac{1}{2\kappa_{11}^2} \int_{r=\infty} \Big\{ \dl \big(e Q^{tr}(\xi_T) \big) 
  + 2 e \xi_T^{[t} \Theta^{r]}(\dl) \Big\} \notag
  \\
  &= \frac{1}{2\kappa_{11}^2} \int_{r=\infty} \Big\{ \delta \big( - 2 e \nabla^{[t} \xi_T^{r]} \big)
  + 4 e \xi_T^{[t} e^{r] a} e^{N b} \dl \om_{N ab} \Big\} \notag
  \\
  &\quad\,
  + \frac{\gamma}{2\kappa_{11}^2} \int_{r=\infty}
  \Big[ \delta \big\{ - 2 e X^{trPQ} \nabla_P \xi_{TQ}
  - 2 e e^t{}_i e^r{}_a ( D_b X^{ijab} - 2 D_b X^{a(ij)b} ) \xi_{Tj} \big\} \notag
  \\
  &\qquad\qquad\qquad
  + 4 e \xi_T^{[t} X^{r]N ab} \dl \om_{N ab} 
  + 4 e \xi_T^{[t} e^{r] a} ( D^b X_{aijb} + D^b X_{ajib} + D^b X_{abij} ) \delta e^{ij} \Big] \notag
  \\
  &= \frac{1}{2\kappa_{11}^2} \int_{r=\infty} \Big\{ \delta \big( - 2 e \nabla^{[t} \xi_T^{r]} \big)
  + 2 e \xi_T^{[t} g^{r]P} g^{NQ} 
  (\nabla_Q \delta g_{PN} - \nabla_P \delta g_{QN}) \Big\}. \label{eq:Wmass}
\end{alignat}
Since each non zero component of the Riemann tensor behaves like $R_{abcd} \sim \mathcal{O}(\frac{1}{r^9})$
asymptotically, in the above we dropped terms which depend on $X_{abcd}$.
The last equation is equivalent to the variation of the ADM mass\cite{IW}.
Therefore the mass is simply given by the ADM mass formula.

When $\xi$ is chosen as an asymptotic translation along $z$ direction $\xi_Z$, it satisfies
$\xi_Z^t \Theta^r(\dl) - \xi_Z^r \Theta^t(\dl) = 0$, and the charge is given by
\begin{alignat}{3}
  Q &= \frac{1}{2\kappa_{11}^2} \int_{r=\infty} e Q^{tr}(\xi_Z) \notag
  \\
  &= \frac{1}{2\kappa_{11}^2} \int_{r=\infty} \big( - 2 e \nabla^{[t} \xi_Z^{r]} \big) \notag
  \\
  &\quad\,
  + \frac{\gamma}{2\kappa_{11}^2} \int_{r=\infty}
  \big\{ - 2 e X^{trPQ} \nabla_P \xi_{ZQ}
  - 2 e e^t{}_i e^r{}_a ( D_b X^{ijab} - 2 D_b X^{a(ij)b} ) \xi_{Zj} \big\} \notag
  \\
  &= \frac{1}{2\kappa_{10}^2} \int_{r=\infty} \sqrt{-g}\, G^{tr} . \label{eq:Wcharge}
\end{alignat}
As explained in the mass formula, terms which depend on $X_{abcd}$ are dropped,
and the dimensional reduction (\ref{eq:dimred}) is used in the last step.
Thus the R-R charge is given by the integral of the R-R flux as usual.

Finally we choose the Killing vector as $\xi = \xi_T + \Omega \xi_Z$,
which becomes zero at the bifurcate horizon $\Sigma$.
Then $\delta_\xi e^M{}_a = \delta_\xi \omega_N{}^{ab} = 0$, and
the symplectic current and the variation of the current also vanish. 
Furthermore, if the variation of the fields, such as $\delta e^M{}_a,$ 
satisfy linearized equations of motion, the eq.~(\ref{eq:varJ2}) is simplified as
\begin{alignat}{3}
  0 &= \frac{1}{2\kappa_{11}^2} \pa_N \big\{ \dl \big(e Q^{MN}(\xi) \big) 
  + e \big( \xi^M \Theta^N(\dl) - \xi^N \Theta^M(\dl) \big) \big\}.
\end{alignat}
By integrating the above equation over 10 dimensional asymptotically flat space with the horizon, 
we derive the first law of the black hole,
\begin{alignat}{3}
  \delta M + \Omega \delta Q = \frac{\kappa}{2\pi} \delta S.
\end{alignat}
Here $\kappa$ is the surface gravity which is given by $\nabla_M \xi_N = \kappa N_{MN}$ 
at the bifurcate horizon $\Sigma$,
and $N_{MN}$ is an antisymmetric tensor which is binormal to the bifurcate horizon $\Sigma$.
$S$ corresponds to the entropy and is given by
\begin{alignat}{3}
  S = - \frac{2\pi}{2\kappa_{11}^2} \int_Q \sqrt{h} 
  (g^{MP} g^{NQ} + \gamma X^{MNPQ}) N_{MN} N_{PQ}.
\end{alignat}
$\sqrt{h}$ is a volume factor of $\Sigma$.

\section{Near Horizon Limit of the Internal Energy}

The internal energy $E$ of the quantum black 0-brane is defined by $E = M - Q$.
From the eqs.~(\ref{eq:mass}) and (\ref{eq:charge}), the internal energy is written as
\begin{alignat}{3}
  E &= \frac{V_{S^8}}{2\kappa_{10}^2} \big( 8 r_-^7 \alpha^7 + 7 r_-^7 - 7 r_-^7 \sqrt{1+\alpha^7}
  - 16865280 \gamma \, r_- \alpha \big). \label{eq:E}
\end{alignat}
Below we consider the near horizon limit of the internal energy, which was first derived in ref.~\cite{H2}
by using the black hole thermodynamics.

Since we will express $E$ in terms of the temperature in the near horizon limit, 
first we need to derive the relation between parameters in the solution and the temperature.
The location of the event horizon $r=r_H$ is obtained by solving $F_1(r_H) = 0$ and modified
up to the linear order of $\gamma$ as
\begin{alignat}{3}
  r_H = r_- \alpha - \frac{\gamma}{7 r_-^5 \alpha^5} f_1(1). \label{eq:hor}
\end{alignat}
The Hawking temperature of the black 0-brane is given by
\begin{alignat}{3}
  T &= \frac{1}{4\pi} H_1^{-\frac{1}{2}} \frac{d F_1}{dr} \Big|_{r_H} \notag
  \\
  &= \frac{7 r_-^\frac{5}{2} \alpha^\frac{5}{2}}{4\pi r_-^\frac{7}{2} \sqrt{1+\alpha^7}}
  \Big[ 1 + \frac{\gamma}{r_-^6 \alpha^6} 
  \Big\{ \Big( \frac{8}{7} - \frac{1}{2(1+\alpha^7)} \Big) f_1(1)
  + \frac{1}{7} f_1'(1) - \frac{1}{2(1+\alpha^7)}h_1(1) \Big\} \Big], \label{eq:tmp}
\end{alignat}
up to the linear order of $\gamma$.
Now let us take the near horizon limit,
\begin{alignat}{3}
  \alpha \to 0, \;\; \frac{r_\pm^7}{\ell_s^{10}} \to (2\pi)^4 15\pi \lambda
  \quad \text{with} \quad U_0 = \frac{r_- \alpha}{\ell_s^2}, \;\;
  \frac{\gamma}{r_-^6 \alpha^6} = \frac{\pi^6 \lambda^2}{2^{7}3^2 N^2 U_0^6}
  \quad \text{fixed.} \label{eq:nhl}
\end{alignat}
Then the temperature (\ref{eq:tmp}) approaches to
\begin{alignat}{3}
  &\tilde{T} = a_1 \tilde{U}_0^\frac{5}{2}
  \big( 1 + \epsilon a_2 \tilde{U}_0^{-6} \big), \label{eq:tmp2}
\end{alignat}
where
\begin{alignat}{3}
  &a_1 = \frac{7}{16\pi^3 \sqrt{15\pi}} \sim 2.06 \times 10^{-3}, \notag
  \\
  &a_2 = \frac{9}{14} f_1(1) + \frac{1}{7} f_1'(1) - \frac{1}{2}h_1(1) \sim - 4.02 \times 10^5, \label{eq:nv}
  \\
  &\epsilon = \frac{\pi^6}{2^{7}3^2 N^2} \sim \frac{0.835}{N^2}. \notag
\end{alignat}
Here dimensional quantities $\tilde{T} \equiv T/\lambda^\frac{1}{3}$ and
$\tilde{U}_0 \equiv U_0/\lambda^\frac{1}{3}$ are introduced to make expressions simple.
From the eq.~(\ref{eq:tmp2}), it is possible to express $\tilde{U}_0$ in terms of $\tilde{T}$ as
\begin{alignat}{3}
  \tilde{U}_0 = a_1^{-\frac{2}{5}} \tilde{T}^\frac{2}{5} 
  \Big( 1 - \frac{2}{5} \epsilon a_1^\frac{12}{5} a_2 \tilde{T}^{-\frac{12}{5}} \Big). \label{eq:U0tilde}
\end{alignat}
By using this relation, the near horizon limit of the internal energy is expressed as a function 
of the temperature.

Finally let us evaluate the near horizon limit of the internal energy.
By taking the near horizon limit (\ref{eq:nhl}) and using 
$2\kappa_{10}^2 = (2\pi)^7 \ell_s^8 g_s^2 = \frac{(2\pi)^{11} \ell_s^{14} \lambda^2}{N^2}$,
$V_{S^8} = \frac{2(2\pi)^4}{7 \cdot 15}$ and the eq.~(\ref{eq:U0tilde}), 
it is calculated as follows.
\begin{alignat}{3}
  E &= \frac{2N^2}{(2\pi)^{7} 105 \ell_s^{14} \lambda^2} 
  \Big( \frac{9}{2} U_0^7 \ell_s^{14} - 16865280 \epsilon \lambda^2 U_0 \ell_s^{14} \Big) \notag
  \\
  &= \frac{9N^2 \lambda^{\frac{1}{3}}}{(2\pi)^{7} 105} 
  \big( \tilde{U}_0^7 - 3747840 \epsilon \tilde{U}_0 \big) \notag
  \\
  &= \frac{9N^2 \lambda^{\frac{1}{3}}}{(2\pi)^{7} 105} 
  \Big\{ a_1^{-\frac{14}{5}} \tilde{T}^{\frac{14}{5}} 
  - \epsilon \Big( \frac{14}{5} a_2 + 3747840 \Big) a_1^{-\frac{2}{5}} \tilde{T}^{\frac{2}{5}} \Big\} .
\end{alignat}
By inserting numerical values of (\ref{eq:nv}),
the dimensionless internal energy of the quantum black 0-brane $\tilde{E} = E/\lambda^\frac{1}{3}$ 
is written as
\begin{alignat}{3}
  \frac{\tilde{E}}{N^2} &= 7.41 \tilde{T}^\frac{14}{5} - \frac{5.77}{N^2} \tilde{T}^{\frac{2}{5}}. 
  \label{eq:Enhl}
\end{alignat}
This is exactly the same as the result derived in ref.~\cite{H2}.
The above relation is also reproduced by the numerical simulation from the dual gauge theory\cite{HHIN},
which strongly supports the gauge gravity duality at the quantum level.

\end{document}